\documentclass[aps,a4paper,preprint]{revtex4}
\usepackage{epsfig}
\usepackage{graphicx}
\usepackage{color}
\usepackage[english]{babel}
\usepackage{fancybox}
\usepackage{amsmath}
\usepackage{amsthm}
\usepackage{eucal}
\usepackage{amssymb}
\usepackage{mathrsfs}
\usepackage{mathtools}
\usepackage{natbib}
\def \beq {\begin{equation}}
\def \eeq {\end{equation}}
\def \beqn {\begin{eqnarray}}
\def \eeqn {\end{eqnarray}}

\def \la {\lambda}

\def \ep {\epsilon}

\def \pr {\partial}

\def \beq { \begin{equation}}

\def \eeq {\end{equation}}

\def \ll {\langle}

\def \rr {\rangle}

\def \st {S^1_\beta \times \mathbb{R}^4_\Omega}

\def \at {\biggl{\vert}}
\begin{document}

\title{ Instanton-torus knot duality in 5d SQED and $SU(2)$ SQCD}

\author{A. Gorsky}
\affiliation{Institute for Information Transmission Problems, B.Karetnyi 19, Moscow, Russia.}
\affiliation{Moscow Institute of Physics and Technology, Dolgoprudny 141700, Russia.}

\begin{abstract}
We briefly review the instanton-torus knot duality found  in 5d SUSY gauge theories with
one compact dimension. The 
fermion condensate  turn out to be the generating function for the torus
knot polynomials colored by the fundamental representation.

\end{abstract}

\maketitle


\section{Introduction}

The field theory derivation of the knot invariants was initiated long time ago
\cite{wittenold} when the Jones polynomial was identified as the vev of Wilson
loop along the knot in the $SU(2)$ 3d Chern-Simons(CS) topological theory. Similar
consideration of $SU(N)$ CS theory yields the HOMFLY polynomials of knot colored 
by the arbitrary representation.
The CS approach  has been embedded into
the topological string framework in \cite{ov,labastida} where the knot is realized via
the intersection of branes in the Calabi-Yau 3-manifold. The invention of the
Khovanov homologies allowed to introduce the generalization of the HOMFLY
knot polynomials - superpolynomials \cite{dgr} .
For instance, the Jones polynomial can be expressed in terms of the Khovanov
homologies $H^{i,j}$ as
\beq
J(q)=\sum_{i,j} (-1)^i q^j dim H^{i,j}
\eeq
The refinement of the CS partition function
which  allows to evaluate the torus knot superpolynomials has been suggested in \cite{agash}
however the refined CS was defined via matrix model and its explicit  field theory representation
is still questionable. The candidate for this field theory has been discussed recently  in \cite{gukovcs}.

The Wilson loop approach is not the only way to recognize the knot
polynomials in the field theory. It was shown in \cite{gsv} that the knot polynomials count the multiplicities
of the BPS states in SUSY YM theories and can be considered as the refined BPS index depending on the set of
chemical potentials which serve as the generating parameters. 
A bit schematically this approach can be summarized in the representation of the superpolynomal for the knot K as 
\beq
P_K(a,q,t)=\sum_{ijk}c_{ijk} a^iq^jt^k
\eeq
where the three gradings correspond to three equivariant parameters for the global
symmetries. The
numbers $c_{ijk}$ count the dimensions of the corresponding spaces of BPS states.

Later the subject got twisted in one more direction and it was argued in \cite{gukov3d} that
via 3d/3d duality the knot invariants can be used for the classification of 3d SUSY
gauge theories. The particular knot at one side of the correspondence fixes the matter
content in the particular dual theory and the knot polynomials serve as some index. 
More recently there was attempt \cite{wittennew} to relate the refined
knot invariants to the instanton counting lifting the CS theory to the topologically
twisted $N=4$ $D=4$ SYM theory  and its $D=5$ counterpart. The nice recent reviews summarizing the
physical and mathematical aspects of the knot homologies   can be found in
\cite{gukovrev} and \cite{nawata}. 

Here we shall briefly review new reincarnation of the knot polynomials found in \cite{gm,gmn}
based on the observation made in \cite{bgn}. The new place for the knot invariants
is the evaluation of the fermion condensates in Omega-deformed 5D 
SQED and $SU(2)$ SQCD.
In an interesting way this novel approach glues together all previously considered viewpoints
concerning the roles of knot polynomials and add new flavor to this issue. The motivation for this new
approach goes as follows. It was argued in \cite{gor10} that bottom row (a=0) of
superpolynomial  for the $(n,n+1)$ torus knots is expressed in terms of the $(q,t)$ deformed Catalan numbers
$C_n(q,t)$
which  are intimately related with the particular equivariant integrals over the
moduli space of centered points on $C^2$ \cite{haiman}. On the other hand 
the evaluation of the HOMFLY polynomials of generic torus knots 
in terms of $Hilb^n(C^2)$ has been developed in \cite{os, ors, gors, vafashende}. It has been
extended to the evaluation of superpolynomial in \cite{gorneg}.

Hence it is natural to suggest that 
in the physical language we are evaluating 
vev of some observable in the K-theory of centered abelian instantons in  5d  gauge theory with
some matter content. The problem concerns the identification
of the particular gauge theory and the particular observable which does the job. The answers to these
questions have been found in \cite{gm,gmn}. 

It turned out that the proper field theory  is the Omega-deformed 5d SQED or $SU(2)$ SQCD with some number
of flavors on $R^4\times S^1$ supplemented in some cases by 5d CS term. The proper observables 
are the  derivatives of the Nekrasov partition functions \cite{nekrasov} with respect to the masses 
of the fundamentals. A bit loosely it could be considered as IR counterpart of the 
fermion condensate defined in the UV.
The two integers characterizing $(n,m)$ torus knot were identified as instanton and electric quantum numbers and
summation over  quantum numbers of BPS particles necessary for evaluation of observable implies 
the summation over the torus  knot polynomials with all $(n,m)$. More precisely the derivative of the Nekrasov partition
function is nothing but the generating function for the torus knot polynomials. The three generating
parameters in the superpolynomial $P_{n,m}(a,q,t)$  of $(n,m)$ torus knot  are 
identified with two equivariant parameters of the Omega-background and 
the mass of the matter in antifundamental representation.

\section{Superpolynomial of $(n,nk+1)$ torus knots}

Recall that the torus knots in $S^3$ can be obtained as follows. The $S^3$ manifold  can 
be represented as two solid tori glued together with the help of S-matrix 
changing the cycles. The boundary of one torus is equipped with the curve 
with winding numbers $(n,m)$ over the cycles. Therefore we can present
the torus knot schematically as $<n,m|S|0.0>$ where $|0,0>$ is the state of pure solid torus. The series $(n,nk+1)$ plays
the special role since the corresponding knots can be obtained from the 
unknot $(n,1)$  in a simple manner. We start with
$(n,1)$ unknot and take into account that due to the Witten's effect the 
instanton with topological charge $n_I$  acquires the electric charge $n_Ik$ where
$k$ is the level of 5d CS term. 
For the $(n,nk+1)$ series  we are able to identify  in physical terms the
superpolynomials depending on three generating parameters $P_{n,nk+1} (a,q,t)$.    
It was shown in \cite{gm,gmn} that the $T_{n,nk+1}$ superpolynomials
are encoded in the UV properties of the condensate of the massless flavor in
the 5d SQED  with one compact dimension  and 5d CS term at level $k$.

To explain the physical picture it is necessary to remind the spectrum of BPS particles
in the 5d SQCD. The corresponding central charge involves the quantum numbers corresponding
to the instantons, W-bosons and fundamentals (for SU(2)) \cite{seiberg}
\beq
Z= \frac{1}{g^2} n_I + n_e a + \sum_i n_{f_i} m_{f_i}
\eeq
where g is coupling constant and a is vev of the real adjoint scalar.
The instantons in 5d theory are particles which carry the charge corresponding to the conserved topological current
\beq
J= * TrF\wedge F
\eeq
If we add to the action the CS term
\beq
S_{CS} = k \int TrA\wedge F\wedge F
\eeq
it implies
that the instanton  particle with charge $n_I$ carries the electric charge $n_Ik$ and the central charge
can be written as 
\beq
Z=(n_e +kn_I)a + \frac{1}{g^2}n_ I  + \sum_i n_{f_i} m_{f_i}
\eeq
Generically a particle carries  quantum numbers $(n_I, n_e, n_f)$. The
specific wall crossing phenomena in the BPS spectrum  have been investigated 
in \cite{hanany,vonk}. The one loop contribution to the 5d effective action involving
all BPS states reproduces the full 4d effective actions upon reduction \cite{Nek5d}.

In \cite{gm} the new instanton-torus knot duality has been formulated for the
Omega-deformed  5d SUSY QED with $N_F=3$ on $\st$ with the Chern-Simons term at level $k$.
It has been proved that the second derivative of the Nekrasov instanton
partition function with respect to the masses of the hypermultiplets  is the
generating function for the superpolynomials of the torus $T_{n,nk+1}$ knots
where $n$ - is the instanton charge.

\beq
\label{super}
\left.\frac{e^{\beta M}}{(1+A)\beta^2}\frac{d^2 Z_{nek}(q,t, m_f,M,m_a,Q,k)}{dM\, dm_f}\right|_{m_f\to 0,
M\to \infty} \medskip \\ = \sum_n Q^n(tq)^{n/2} P_{n,nk+1}(q,t,A)
\eeq
where $m_a,m_f,M$ are masses of three hypermultiplets in antifundamental ($m_a$) and 
fundamental representations ($m_f,M$)
and $Q$ is the counting parameter for the instantons. The mapping 
between the parameters at the lhs and rhs goes as follows

\begin{eqnarray}
t=\exp(-\beta \epsilon_1) \\
q=\exp(-\beta \epsilon_2) \\
A=-\exp(\beta m_a) \\
Q=\exp(-\beta/g^2)
\end{eqnarray}
where $\beta$ corresponds to the radius of $S^1$ and $\epsilon_1, \epsilon_2$
are two equivariant parameters of the Omega-background which physically
mean two independent angular velocities in $R^4$.

The selection of two special masses  provides the reduction of the generic double sum 
over quantum numbers  to the
summation over the  instanton number without the summation over the electric charges. 
One could say that we select the particular observable in the gauge theory 
with the fixed value of the electric charge. Indeed  it is possible to trade the regulator field with mass M
to the insertion of the particular operator $\exp(\beta \phi)$,  where $\phi$ is the
scalar in adjoint, inserted 
at some point in $R^4$ which is the particular example of qq- character \cite{qq}. This realization of the generating function for the
superpolynomials explains why the answer
deals with the instantons centered around the point with the operator insertion. 

Hence we can claim that from the 5d theory viewpoint the refined knot 
invariants are sensitive to the interference of  two independent 
rotations in $R^4$  and the regulator UV scale M. Since the abelian 
instantons are centered  we could say a bit loosely that the refined
knot invariants govern the renormalization effects from the point-like
instantons or sheaves on Hilbert scheme of $C^2$ in more mathematical terms. The rich life at point in $R^4$ can be also recognized
from the CY side  since in M theory the instantons actually are 
M2 branes extended in CY manifold. 
To conclude this Section  we  mention  that it is worth to think that the information about the knot superpolynomial 
is encoded  in the non-perturbative UV properties of the fermionic condensate of the massless flavor.

\section{The HOMFLY polynomials  for generic $(n,m)$ torus knots from $SU(2)$ SQCD}
\subsection{$N_f=2$ theory with Lagrangian brane}

In this Section we shall consider the unrefined case and corresponding HOMFLY
polynomials. We switched off the second independent parameter of the background 
however as a bonus  are able to represent the fermion condensate as the double series over two quantum numbers
and therefore over all torus knots.
To this aim instead of the insertion of the particular operator in $N_f=2$ theory we can consider  the
$N_f=2$ $SU(2)$ SQCD  supplemented by the Lagrangian brane with zero framing with some value of Fayet-Iliopulos(FI) parameter $z$.
To get the HOMFLY polynomials we make two step procedure. First, we consider the decoupling
limit $1/g^2 \rightarrow \infty$ in $SU(2)$ theory when it effectively decouples into the product of
two $U(1)$ theories and pure 4d instantons decouple. However due to the additional Lagrangian
brane we have the  FI parameter which counts the instantons on the Lagrangian brane \cite{holland}.
Considering the derivative of the Nekrasov partition function in this case with respect to mass
and expanding it into the double series $z^m Q^n$ we obtain the HOMFLY polynomials
of the generic $(n,m)$ knots as the coefficients of the expansion. 

\beq
\ll \tilde{\psi} \psi \rr_{LB}=\frac{\pr Z_{inst}}{\pr m_f} \biggl{\vert}_{m_f=0}= \sum_{n,m} Q_c^n z^m P_{n,nk+m}(A,q,t)
\eeq
\begin{eqnarray}
\label{new}
P(A,q,t)_{n,nk+m}= \\ \nonumber
 \sum_{\la : |\la|=n }  \cfrac{t^{(k+1) \sum l} q^{(k+1) \sum a}(1-t)\prod^{0,0}(1+A q^{-a'}t^{-l'})\prod^{0,0}(1-q^{a'}t^{l'})}{\prod(q^a-t^{l+1})\prod(t^l-q^{a+1})} \times Coef_{z^m} M(z) 
\end{eqnarray}
where $M(z)$ is the contribution from the Lagrangian brane with zero framing:
\beq
M(z) = \prod_{j=1}^{l(\la)} \frac{1-z t^{j-1} q^{\la_j}}{1-z t^{j-1}} 
\eeq

In this case the parameter $z$ counts 
the 2d instantons while the parameter Q equals to $\exp(\beta a)$ and counts the number
of W- bosons in the decoupling perturbative limit of $SU(2)$ theory. In this approach we can say
that HOMFLY polynomials provide the entropic factor in the condensate in the sector with
the particular defect. 

\subsection{$SU(2)$ $N_f=4$ SQCD }

The most effective way to evaluate the effective action in 5d gauge theory is the
use of the refined vertex \cite{vertex}. The counting of the torus knot invariants
along this way can be found in \cite{klemm,einard}. We can use some useful 
properties of the toric diagrams for $SU(2)$ SQCD 
to embed the abelian theory  into the SU(2) with $N_f=4$. Remind that the abelian
5d theory is engineered by $O(-1)\times O(-1)\rightarrow CP_1$ CY manifold. The
5d CS term adds the nontrivial framing while each flavor corresponds to blow-up
of one point.

To get the $SU(2)$ theory with $N_f=4$ it is necessary to perform  two transformations
of the toric diagrams \cite{gmn} trading the Lagrangian brane for the additional flavor.
We end up with two flavors in fundamental and two in antifundamental. The corresponding
masses are $ m_1 \rightarrow 0, \quad m_2= m_a, \exp(\beta m_3) = q\sqrt{qt}, \exp(\beta m_4)= \sqrt{qt}$.
Hence two masses
of fundamentals are fixed by parameters of the $\Omega$-deformation, one mass tends to zero and
one mass is arbitrary. The toric diagram with the corresponding Kahler parameters 
is presented at the Figure.

\begin{figure}[!h]
\begin{center}
\setlength{\unitlength}{1.2cm}
\begin{picture}(7,6)
\linethickness{0.3mm}
\put(2,2){\line(1,0){1}}
\put(3,1){\line(0,1){1}}
\put(3,2){\line(1,1){1}}
\put(4,3){\line(1,0){1}}
\put(4,3){\line(0,1){1}}
\put(2,4){\line(1,0){2}}
\put(4,4){\line(1,1){1}}
\put(5,5){\line(0,1){1}}
\put(5,5){\line(1,0){1}}
\put(2,2){\line(0,1){2}}
\put(2,2){\line(-1,-1){1}}
\put(2,4){\line(-1,1){1}}
\put(1,1){\line(-1,0){1}}
\put(1,1){\line(0,-1){1}}
\put(1,5){\line(-1,0){1}}
\put(1,5){\line(0,1){1}}
\put(2.5,1.7){\makebox{$Q$}}
\put(4.7,4.4){\makebox{$Q_a$}}
\put(4.1,3.5){\makebox{$Q_c$}}
\put(3.7,2.4){\makebox{$Q_f$}}
\put(3,4.2){\makebox{$z$}}
\put(1.6,1.2){\makebox{$\sqrt{q t}$}}
\put(0.7,4.2){\makebox{$q \sqrt{q t}$}}
\end{picture}
\end{center}
\caption{$SU(2)$ theory with four flavors}
\label{fg:su2_m4}
\end{figure}
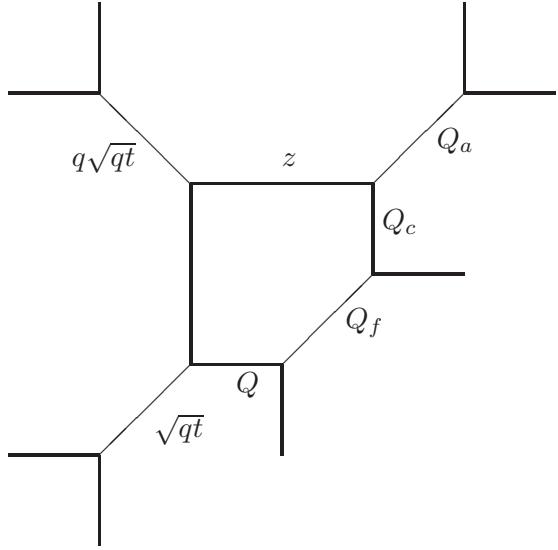

If we 
expand the derivative of the partition function into the double 
series $e^{m \beta a} Q^n$ corresponding to the expansion in the electric and 
4d instanton charges we get the HOMFLY polynomial for the generic torus knots \cite{gmn}. 
No decoupling of 4d instantons occurs since $Q= \exp(\beta g^{-2})$ is finite.
In terms of the toric diagram in IIB picture the instantons correspond to the horizontal D1 
strings while the W-bosons to the vertical F1 strings in the box of the toric diagram. 
The complicated string
network in IIB has to be taken into account for the generic quantum numbers.
The representation of the HOMFLY polynomials  in terms of $SU(2)$ $N_f=4$ SQCD explains their
relation with the q-Liouville conformal blocks via AGT relation \cite{agt,awata} 
which has been observed in \cite{gm}.

\section{Fractional 5d CS term and the Calogero model }

Another approach involves the famous Jones-Rosso formula for the torus knots.
It was shown in \cite{gmn} that Jones-Rosso representation for the HOMFLY polynomial 
corresponds to the $N_f=2$ SQED without the
additional operator insertions  but with the fractional 5d CS term. The Jones -Rosso formula reads as 
\beq
\label{homfl2}
H^{(n,m)}_\square(A,q)=(-1)^{n-1}\frac{1-q^n}{q^n} \sum_{|\la|=n}   q^{(\frac{m}{n}+1)\sum(l-a)}
\frac{\prod^{0,0}(1-q^{l'-a'}) \prod^{0,0}(1+A q^{a'-l'})}{\prod(q^{-l-1}-q^a)(q^{-l}-q^{a+1})} 
\eeq
and fits with the instanton partition function of 5D $\mathcal{N}=1$ U(1) gauge theory on 
$\mathbb{R}^4_\Omega \times S^1_\beta$ in self-dual Omega deformation $\epsilon_1=-\epsilon_2$ with antifundamental
matter of mass $m_a$ and fundamental matter of mass $m_f$ supplemented with fractional CS term:
\beq
\frac{\pr \tilde{Z}^{inst}_n}{\pr m_f} \at_{m_f=0} = (1+A) \beta \sum_{|\la|=n}   q^{(\frac{m}{n}+1)\sum(l-a)}
\frac{\prod^{0,0}(1-q^{l'-a'}) \prod^{0,0}(1+A q^{a'-l'})}{\prod(q^{-l-1}-q^a)(q^{-l}-q^{a+1})} 
\eeq
where the parameters are identified as follows
\beq
q=\exp(-\beta \ep_2),\ A=-\exp(\beta m_a)
\eeq
The generalization of this formula for the superpolynomial of generic $T_{n,m}$ knot has been found in \cite{gorneg}.

In this representation we  obtain the $n$-instanton contribution to the fermionic  condensate
as the HOMFLY invariant of $T_{n,nk+1}$ knot when the fractional 5d CS
term is $k=m+1/n$. Note that the denominator in the CS coupling is equal to the number of instantons. This approach does not allow to get the whole instanton sums but provides the additional 
framework for the evaluation of the separate terms in the instanton expansion of the 
fermion condensate.

Due to the non-vanishing CS term the instanton acquires the electric  charge $k+1$ therefore
in this case we have the system of n instanton dyons. It is known that the 5d CS term induces
the connection on the instanton moduli space which results in our case to the
interacting instanton dyons in $R^4$. Remarkably it turns out that the system of the interacting 
instanton dyons in $R^4$ is described by the integrable n-particle Calogero model with the attractive
interaction. The coupling constant is determined by the coefficient in front of the 5d CS term. 
The HOMFLY polynomial of $T_{n,m}$ knots appears in the spectrum of the Calogero model with the  
rational $m/n$ coupling constant in an interesting way as the
weighted degeneracy of the $E=0$ states \cite{gordaha}. More algebraically HOMFLY polynomials can 
be identified with the twisted character of the finite-dimensional representation 
of DAHA which exists at rational coupling constant in the Calogero model \cite{berest}.
Physically the situation corresponds to the "`falling to the center"' problem and reflects
the centering of the instantons. Let us emphasize that the Calogero dynamics takes place
in the $R^4$ while the torus knots are pictured inside the CY space by the membrane instantons
which are actual degrees of freedom in the Calogero model.

This representation can be potentially  useful for the description of the FQHE in (4+1) dimensions.
The Calogero coupling constant corresponds to the filling fraction and it is well-known that the
Hall conductivity $\sigma_H$ is identified with the coefficient in front of the 5d CS term \cite{fqhe}.
It seems that the  interacting instanton dyons are some analogue of the 
composite fermions in the (2+1) FQHE. The changing of the 5d  CS level corresponds to the 
transition between two plateau which has on the other hand the interesting geometrical 
interpretation as the cut-and-join operation \cite{dunin}.

\section{Where the duality comes from?}

Let us briefly summarize the qualitative physical picture behind the duality found. The most 
unusual features are the necessity to sum over all torus knots to get the vev of physical 
observable and the relation between the mass of the field in fundamental representation and the
rank of the gauge group involved into the 3d CS derivation of the HOMFLY invariants. Actually
these features are related: the instantons and instanton dyons are represented by the wrapped M2 branes while
the mass of the fundamental being the Kahler modulus of $CP_1$ gets traded for the rank of the 
corresponding gauge group in 3d CS theory or for the number of M5 branes upon the inverse geometrical transition.
The M2 branes draw the torus knots on the M5 branes representing the fundamental 
(or anfifundamental) matter inside the CY manifold. The necessity to sum over the all knots is now clear. The similar
picture can be seen also in IIB model where the relation with the toric diagrams for CY manifold makes the 
underlying geometry even more transparent \cite{gmn} and the BPS states are represented by the
string web.

The knots are placed in CY space and 
polynomials of $T_{n,m}$ knots in CY space count the multiplicity of the particles with $(n,m)$ charges
in the 5d gauge theory.  The most subtle point concerns the choice of the observable however 
upon this choice is done the knot HOMFLY polynomials can be equally  evaluated in terms of the spectrum of Calogero 
model in physical space-time and describes the peculiarities of  the "`falling to the center"' phenomena for the 
instanton dyons.

To explain the interplay between the mass of the antifundamental and the rank of 3d CS group consider 
first one-loop effective action in  QED in  constant  external electric and magnetic fields. 
In a self-dual background field the effective action can 
be identified as the topological string at $T^*S^3$ or equivalently  $SU(N)$ 3d CS at $S^3$ when the rank 
of the  group appears to be the ratio of the fermion mass and the external field $N \propto \frac{m^2}{eE}$
\cite{gorly}. This is the toy example of the inverse geometrical transition  with the 
mass dependent rank of the group in the 3d CS theory.

Assume now that we have the second fermion loop in the same external field probably of the
different  flavor. We take the derivative of the second loop with respect to the mass 
which yields the insertion of  fermion bilinear.
At the next step we assume that there is the web of interacting particles   of two types 
between the loop with operator insertion and the first loop. 
They can braid providing the torus knots $T_{n,m}$ if we have $n$ propagating
particles of one type and $m$ propagating particles of another type. Since we have prepared 
3d CS theory in CY space from the first loop in the external field the ends of the propagating 
particles picture the torus knot in $S^3$ inside CY. From the viewpoint of the second loop with
the operator inserted we evaluate the contribution to the condensate from the "tadpole" connected
to the loop by some web involving particles of two types. The knot invariants count the entropy 
of the web 
with fixed two quantum numbers which are attached to the loop. Equivalently,
it can be thought as the particular entropic factor in the fermion  condensate.
To some extend this picture can be considered as the generalization of the Schwinger-type
evaluation of the BPS multiplicities in \cite{gopakumar}.

In our case we have the loop of the antifundamental in the external graviphoton field (at least at
weak field)  and the
loop of the fundamental in the same external field with the inserted bilinear operator.
 Due to the inverse geometric transition the loop of antifundamental provides the $SU(N)$ 3d CS action 
in CY when the rank of the group is $N \propto \frac{m_a}{\epsilon}$ similar to the
QED case above. The insertion of the fermion bilinear and the loop of antifundamental 
are connected by the instanton-W-boson web
with electric and instanton charges $(n,m)$ which pictures the torus knot at the
antifundamental side. From the viewpoint of the fundamental we evaluate the condensate of
the bilinear in the external field taking into account the tadpole of the antifundamental connected
by the W-boson-instanton web. The configuration of the web has some peculiarities, for instance,
one has to have in mind that instantons are almost sitting at the top of each other in $C^2$.
The multiplicity of the web yields the entropic factor in the condensate.

Our findings suggest that
the role of the polynomials of the torus knots drawn in the "` momentum space"' could be fairly important for the
description of the different vacuum condensates in the field theory. The key phenomena behind
the appearance of the knots (at least torus knots) is the interesting interplay between
the perturbative and non-perturbative contributions to the physical observables. This could
be crucial for the resurgence phenomena in generic QFT.

\section{Conclusion and Acknowledgements}

We have briefly summarized the recent developments concerning the relation
between the instanton counting in 5d SQED and $SU(2)$  SQCD and torus knot invariants. There are
a lot of open questions, some of
them are under investigation now \cite{gmn2}.

I am grateful to K. Bulycheva, S. Nechaev and especially to A. Milekhin and N.Sopenko
for the collaboration on these issues. The results were reported at SCGP Workshop "`Future Prospects for Fundamental Particle Physics and Cosmology", May 4 – 8, 2015, workshop "`Gribov -85"' Chernogolovka , June 2015 and JINR workshop "`Supersymmetries and Quantum Symmetries", August 2015.  The work was supported by grant RFBR-15-02-02092.


\begin{thebibliography}{99}
\bibitem{wittenold} 
  E.~Witten,
  ``Quantum Field Theory and the Jones Polynomial,''
  Commun.\ Math.\ Phys.\  {\bf 121}, 351 (1989).
\bibitem{ov}
  H.~Ooguri and C.~Vafa,
  ``Knot invariants and topological strings,''
  Nucl.\ Phys.\  B {\bf 577}, 419 (2000)
  [arXiv:hep-th/9912123].
	\bibitem{labastida} 
  J.~M.~F.~Labastida and M.~Marino,
  ``Polynomial invariants for torus knots and topological strings,''
  Commun.\ Math.\ Phys.\  {\bf 217}, 423 (2001)
  [hep-th/0004196].

\bibitem{dgr}
 N. Dunfield, S. Gukov and J. Rasmussen,
 "`The Superpolynomial for Knot Homologies"',
 arxiv;math/0505662
	\bibitem{agash} 
  M.~Aganagic and S.~Shakirov,
  ``Knot Homology from Refined Chern-Simons Theory,''
  arXiv:1105.5117 [hep-th].
\bibitem{gukovcs} 
  S.~Gukov and D.~Pei,
  ``Equivariant Verlinde formula from fivebranes and vortices,''
  arXiv:1501.01310 [hep-th].
\bibitem{gsv}
  S.~Gukov, A.~S.~Schwarz and C.~Vafa,
  ``Khovanov-Rozansky homology and topological strings,''
  Lett.\ Math.\ Phys.\  {\bf 74}, 53 (2005)
  [arXiv:hep-th/0412243].
\bibitem{gukov3d} 
  T.~Dimofte, D.~Gaiotto and S.~Gukov,
  ``Gauge Theories Labelled by Three-Manifolds,''
  Commun.\ Math.\ Phys.\  {\bf 325}, 367 (2014)
  [arXiv:1108.4389 [hep-th]].\\
H.~J.~Chung, T.~Dimofte, S.~Gukov and P.~Sulkowski,
  ``3d-3d Correspondence Revisited,''
  arXiv:1405.3663 [hep-th].
\bibitem{wittennew}
  E.~Witten,
  ``Fivebranes and Knots,''
  arXiv:1101.3216 [hep-th].\\
 D.~Gaiotto and E.~Witten,
  ``Knot Invariants from Four-Dimensional Gauge Theory,''
  Adv.\ Theor.\ Math.\ Phys.\  {\bf 16}, no. 3, 935 (2012)
  [arXiv:1106.4789 [hep-th]].
\bibitem{gukovrev} 
  S.~Gukov and I.~Saberi,
  ``Lectures on Knot Homology and Quantum Curves,''
  arXiv:1211.6075 [hep-th].
	\bibitem{nawata} 
  S.~Nawata and A.~Oblomkov,
  ``Lectures on knot homology,''
  arXiv:1510.01795 [math-ph].
\bibitem{gm} 
  A.~Gorsky and A.~Milekhin,
  ``Condensates and instanton – torus knot duality. Hidden Physics at UV scale,''
  Nucl.\ Phys.\ B {\bf 900}, 366 (2015)
  doi:10.1016/j.nuclphysb.2015.09.015
  [arXiv:1412.8455 [hep-th]].
\bibitem{gmn} 
  A.~Gorsky, A.~Milekhin and N.~Sopenko,
  ``The Condensate from Torus Knots,''
  JHEP {\bf 1509}, 102 (2015)
  doi:10.1007/JHEP09(2015)102
  [arXiv:1506.06695 [hep-th]].
\bibitem{bgn} 
 K.~Bulycheva, A.~Gorsky and S.~Nechaev,
  ``Critical behavior in topological ensembles,''
  Phys.\ Rev.\ D {\bf 92}, no. 10, 105006 (2015)
  [arXiv:1409.3350 [hep-th]].
\bibitem{gor10} 
E.~Gorsky,
''q,t-Catalan numbers and knot homology,''
Zeta functions in algebra and geometry, 213-232, Contemp. Math., {\bf 566}, Amer. Math. Soc., Providence, RI, 2012 
[arXiv:1003.0916]  
\bibitem{haiman}
M.~Haiman, "`(q,t) Catalan numbers and the Hilbert scheme"', Discrete Mathematics , 193, (1998),  201 \\
A. M. Garsia, M. Haiman, 
''A remarkable q, t-Catalan sequence and q-Lagrange inversion,'' 
Journal of Algebraic Combinatorics,  {\bf5(3)}, 191-244.
\bibitem{os}
A. Oblomkov and  V. Shende
"`The Hilbert scheme of a plane curve singularity and the HOMFLY polynomial of its link"'
Duke Math. J. Volume 161, Number 7 (2012), 1277-1303 
\bibitem{ors}
A.~Oblomkov, J.~Rasmussen, V.~Shende,
''The Hilbert scheme of a plane curve singularity and the HOMFLY polynomial of its link,''
Duke Mathematical Journal, 03/2010  
[arXiv:1201.2115].
\bibitem{gors}
E. Gorsky, A. Oblomkov, J. Rasmussen and V. Shende
"Torus knots and DAHA representations"
arXiv; 1207.4523 ,Duke Math. J. 163, no. 14 (2014), 2709-2794
	\bibitem{vafashende}
 D.~E.~Diaconescu, V.~Shende and C.~Vafa,
  ``Large N duality, lagrangian cycles, and algebraic knots,''
  Commun.\ Math.\ Phys.\  {\bf 319}, 813 (2013)
  doi:10.1007/s00220-012-1563-3
  [arXiv:1111.6533 [hep-th]].

\bibitem{gordaha}
E.~Gorsky, " Arc spaces and DAHA representations",
Selecta Mathematica 19 (2013) 125, [arxiv:1110.1674]

\bibitem{gorneg} 
  E.~Gorsky and A.~Negut,
  ``Refined knot invariants and Hilbert schemes,''
  arXiv:1304.3328 [math.RT].
	
\bibitem{nekrasov} 
  N.~A.~Nekrasov,
  ``Seiberg-Witten prepotential from instanton counting,''
  Adv.\ Theor.\ Math.\ Phys.\  {\bf 7}, 831 (2004)
  [hep-th/0206161].\\
 N.~Nekrasov and A.~Okounkov,
  ``Seiberg-Witten theory and random partitions,''
  hep-th/0306238.	
	
	
	\bibitem{qq} 
  N.~Nekrasov,
  ``Non-Perturbative Schwinger-Dyson Equations: From BPS/CFT Correspondence to the Novel Symmetries of Quantum Field Theory,'' 
   ``Pomeranchuk 100'', World Scientific 2014.\\
	 N.~Nekrasov,
  ``BPS/CFT correspondence: non-perturbative Dyson-Schwinger equations and qq-characters,''
  arXiv:1512.05388 [hep-th].
	

	\bibitem{seiberg} 
  N.~Seiberg,
  ``Five-dimensional SUSY field theories, nontrivial fixed points and string dynamics,''
  Phys.\ Lett.\ B {\bf 388}, 753 (1996)
  [hep-th/9608111].\\
 D.~R.~Morrison and N.~Seiberg,
  ``Extremal transitions and five-dimensional supersymmetric field theories,''
  Nucl.\ Phys.\ B {\bf 483}, 229 (1997)
  [hep-th/9609070].
\bibitem{hanany}
	O.~Aharony, A.~Hanany and B.~Kol,
  ``Webs of (p,q) five-branes, five-dimensional field theories and grid diagrams,''
  JHEP {\bf 9801}, 002 (1998)
  doi:10.1088/1126-6708/1998/01/002
  [hep-th/9710116].
	\bibitem{vonk}
	 E.~P.~Verlinde and M.~Vonk,
  ``String networks and supersheets,''
  hep-th/0301028.
	\bibitem{Nek5d}
A.~E.~Lawrence and N.~Nekrasov,
  ``Instanton sums and five-dimensional gauge theories,''
  Nucl.\ Phys.\ B {\bf 513}, 239 (1998)
  [hep-th/9706025].


	
	
	\bibitem{holland} 
  T.~Dimofte, S.~Gukov and L.~Hollands,
  ``Vortex Counting and Lagrangian 3-manifolds,''
  Lett.\ Math.\ Phys.\  {\bf 98}, 225 (2011)
  [arXiv:1006.0977 [hep-th]].



\bibitem{vertex} 
  A.~Iqbal, C.~Kozcaz and C.~Vafa,
 ``The Refined topological vertex,''
  JHEP {\bf 0910}, 069 (2009)
  [hep-th/0701156].

  





\bibitem{einard}
  A.~Brini, B.~Eynard and M.~Marino,
  ``Torus knots and mirror symmetry,''
  arXiv:1105.2012 [hep-th].







\bibitem{dunin} 
  P.~Dunin-Barkowski, A.~Mironov, A.~Morozov, A.~Sleptsov and A.~Smirnov,
  ``Superpolynomials for toric knots from evolution induced by cut-and-join operators,''
  JHEP {\bf 03}, 021 (2013)
  [JHEP {\bf 1303}, 021 (2013)]
  [arXiv:1106.4305 [hep-th]].

	\bibitem{fqhe}
	J. Hu and S. Zhang  ,"'Collective excitations at the boundary of a 4D Quantum Hall droplet"',
cond-mat 0112432
	

\bibitem{agt} 
  L.~F.~Alday, D.~Gaiotto and Y.~Tachikawa,
  ``Liouville Correlation Functions from Four-dimensional Gauge Theories,''
  Lett.\ Math.\ Phys.\  {\bf 91}, 167 (2010)
  [arXiv:0906.3219 [hep-th]].
	\bibitem{awata} 
  H.~Awata and Y.~Yamada,
  ``Five-dimensional AGT Conjecture and the Deformed Virasoro Algebra,''
  JHEP {\bf 1001}, 125 (2010)
  [arXiv:0910.4431 [hep-th]].
	

\bibitem{klemm} 
  H.~Jockers, A.~Klemm and M.~Soroush,
  ``Torus Knots and the Topological Vertex,''
  Lett.\ Math.\ Phys.\  {\bf 104}, 953 (2014)
  [arXiv:1212.0321 [hep-th]].
\bibitem{berest}
Y. Berest, P. Etingof, V. Ginzburg, 
''Finite-dimensional representations of rational Cherednik algebras,'' 
Int. Math. Res. Not. 2003, no. {\bf 19}, 1053�1088
[arXiv:math/0208138].


\bibitem{gorly}
	 A.~Gorsky and V.~Lysov,
  ``From effective actions to the background geometry,''
  Nucl.\ Phys.\ B {\bf 718}, 293 (2005)
  doi:10.1016/j.nuclphysb.2005.04.020
  [hep-th/0411063].
\bibitem{gopakumar} 
  R.~Gopakumar and C.~Vafa,
  ``M theory and topological strings. 2.,''
  hep-th/9812127.
\bibitem{gmn2}
	A. Gorsky, A. Milekhin and N. Sopenko in preparation



\end{thebibliography}
\end{document}